\documentclass[apj,twocolumn]{openjournal}

\usepackage{color}
\usepackage{natbib} 
\usepackage{aas_macros} 
\usepackage{amssymb}
\usepackage{amsmath}
\usepackage{hyperref}	% Hyperlinks
\hypersetup{colorlinks=true,linkcolor=blue,citecolor=blue,filecolor=blue,urlcolor=blue}
\usepackage[caption=false]{subfig}

\shorttitle{SMUDGes Detritus}
\shortauthors{}

\begin{document}

\title{Systematically Measuring Ultra-Diffuse Galaxies. VIII. Misfits, Miscasts, and Miscreants}

%\author[0000-0002-5177-727X]{Dennis Zaritsky}
\author{Dennis Zaritsky}

%[0000-0001-7618-8212]
\author{Richard Donnerstein}

%[0000-0002-7013-4392]
\author{Donghyeon J. Khim}

\affiliation{Steward Observatory, University of Arizona, 933 North Cherry Avenue, Tucson, AZ 85721-0065, USA}

\email{corresponding email: dennis.zaritsky@gmail.com}

%%--------------------------------------- ABSTRACT ---------------------------------------%%

\begin{abstract}
\noindent
We re-examine the 7,070 candidate ultra-diffuse galaxies (UDGs) in the SMUDGes survey and provide classifications based on their visual morphology. Among the more interesting cases, we identify objects along a low surface brightness galaxy merger sequence (ongoing mergers (8) and post-mergers (7)) and a distinct set of dwarf ring galaxies (29). The ring galaxies are hypothesized to be the result of nearly polar-axis collisions, but the responsible companions are undetected. We also highlight objects in the catalog that appear to be tidally affected (66), thereby cautioning that their cataloged parameters may be unreliable. Finally, we identify contaminants of various types in the catalog, leaving 6,553 as viable  undisturbed UDG candidates. We discuss all categories and provide example images of the more interesting ones. 
\end{abstract}


\section{Introduction}

Cataloging may appear to be an unscientific endeavor. One decides what one is looking for; one looks for it; one finds it. However, catalogs are an ever-present part of astronomy, from that produced by Messier \citep{messier} to that produced from the Sloan Digitial Sky Survey \citep{sdss}. Catalogs satisfy a fundamental need by providing samples of objects of interest for subsequent study and, perhaps even more important, by illuminating classification challenges and uncovering the unexpected. 

In our own line of investigation \citep{Zaritsky+2019,Zaritsky+2022,smudges5}, where we focus on compiling and then analyzing a large sample of physically large, low surface brightness galaxies \citep[otherwise known as ultra-diffuse galaxies or UDGs;][]{vanDokkum+2015}, we have faced such classification challenges and, on occasion, been surprised \citep{kite}.
In \cite{smudges5} we discuss the full SMUDGes sample, including the distribution in distance and environment, when those measurements are available. We are in the process of improving our measurement of environment and preparing additional measurements, such as a quantification of the recent star formation histories, which we will presented in upcoming work. 

In this installment in our series of papers describing our effort to Systematically Measure Ultra-Diffuse Galaxies (the SMUDGes survey), we return to our published catalog \citep{smudges5} and look more closely at the objects that may not be quite what we had intended to catalog.
In doing this, we describe and discuss the heterogeneous population of objects that strictly satisfy our candidate UDG criteria (central surface brightness in the $g-$band of 24 mag arcsec$^{-2}$ or fainter and an effective radius, $r_e$, of 5.3 arcsec or greater, as derived from a single 2-D single S\'ersic best-fit model to each detection). We do not focus here on the conversion to physical parameters needed to confirm that a candidate UDG satisfies the UDG criterion of having a physical $r_e \ge$ 1.5 kpc. Rather, we are concerned with cataloged objects that are, in various ways, departures from the envisioned UDG class, consisting of relaxed systems that are described well by a single S\'ersic profile, plus, possibly, a nuclear star cluster at their center \citep[e.g.,][]{lambert,khim}. 

Some of these interlopers may meet the spirit of the selection, but some definitely do not. Although the latter should certainly be excluded from any discussion of UDGs, the inclusion or exclusion of the former is a thornier problem and can affect the subsequent interpretation of the UDG population and our modeling of it. 
We summarize the data in \S\ref{sec:data}; present the results of our classification in \S\ref{sec:results}, including detailed discussions of the misfits, miscasts, and miscreants; discuss implications for the study of UDGs in \S\ref{sec:discussion}; and summarize in \S\ref{sec:summary}. Where needed, we adopt the parameters of a flat $\Lambda$CDM cosmological model \citep{hinshaw} and magnitudes are on the AB system \citep{oke1,oke2}.

\section{Data}
\label{sec:data}

The primary data source for the SMUDGes survey is the southern and northern sky images of the DESI Legacy Survey \citep{Dey+2019}. We described the procedure that we use to identify the low-surface sources that comprise the SMUDGes catalog in a series of papers \citep{Zaritsky+2019,Zaritsky+2021,Zaritsky+2022,smudges5}. As part of \cite{smudges5}, in which we present our completed catalog, we  visually examined each of the 7070 candidates and flagged those that we found to be the most egregious misclassifications. Here, we review the entire list of candidates, evaluate our classification of all, and categorize the objects we identify as problematic. We now go beyond a simple binary flag and provide our best estimate of what the problematic detections might actually be. We will discuss the more interesting categories of SMUDGes, including mergers of low surface brightness galaxies, dwarf ring galaxies, and tidally distorted systems.

Our classifications still depend on visual assessments, often made using low signal-to-noise detections. As such, we aim to err on the side of not flagging a detection, but there are ambiguous cases. Where possible, we complement the Legacy data, with the deeper DR3 HyperSuprime Camera imaging \citep{aihara} that is now also available through the Legacy Survey Browser. These data are limited on the sky, so this deeper, multi-band imaging is available for only 577 of the SMUDGes.

In a few cases, we can turn to spectroscopic redshifts to assess our designations. Our redshift compilation comes from a variety of sources, including any previously compiled by NED and SIMBAD and confirmed matches to within 5.3 arcsec in DESI DR1 \citep{desi}. Finally, we include those redshifts we have obtained as part of related programs \citep{Kadowaki+2017,Kadowaki+21,karunakaran24,ascencio} and from independent dwarf galaxy observations \citep{matlas22,matlas23}. 

We provide a line-by-line matched catalog, Table 1, to the original catalog, where the column titled {\bf Rejected} in the original catalog can now be replaced with our new column {\bf Class}. Using this new table, one can select which categories to retain or reject depending on one's interests. We discuss the meaning of each classification category below and present images as examples when necessary.

\begin{deluxetable}{lr}
\label{tab:class}
\tablewidth{0pt}
\tablecaption{Classifications}
\tablehead{\colhead{Name} & \colhead{Class}}
\startdata
SMDG0000017+325141&G\\
SMDG0000202$-$402435&G\\
SMDG0000334+165424&G\\
SMDG0000453+305356&G\\
SMDG0000473$-$040432&TDW\\
SMDG0000500$-$040035&D\\
SMDG0000588$-$410921&G\\
SMDG0001278$-$553133&G\\
SMDG0001293$-$441647&B
\enddata
\tablenote{First 10 rows of the classification table. Full table available.}
\end{deluxetable}

\section{Results}
\label{sec:results}

\subsection{Misfits}

\medskip
In this category we place SMUDGes that are likely to be diffuse, low surface brightness galaxies, but whose structural properties are  distorted at present, as observed, by tidal forces. To be clear, their measured sizes, axis ratios, and total luminosities may not be the values that they will eventually settle on once the system has time to relax, nor may they reflect their initial values. Including these objects in the study of UDGs complicates the interpretation in a number of ways. For example, a study aiming to understand the three dimensional shape of UDGs by modeling the distribution of projected axis ratios \citep[cf.][]{burkert17} should not include these objects. However, these SMUDGes should not be rejected entirely. They are indeed low surface brightness (LSB) galaxies and will likely settle to become regular LSB galaxies, perhaps even UDGs. They are examples of a stage along at least one distinct LSB/UDG formation pathway that invokes galaxy transformation via interactions \citep[e.g.][]{Carleton+2019}. 

\subsubsection{The Tidally Distorted}

\begin{figure}
\includegraphics[width=0.482\textwidth]{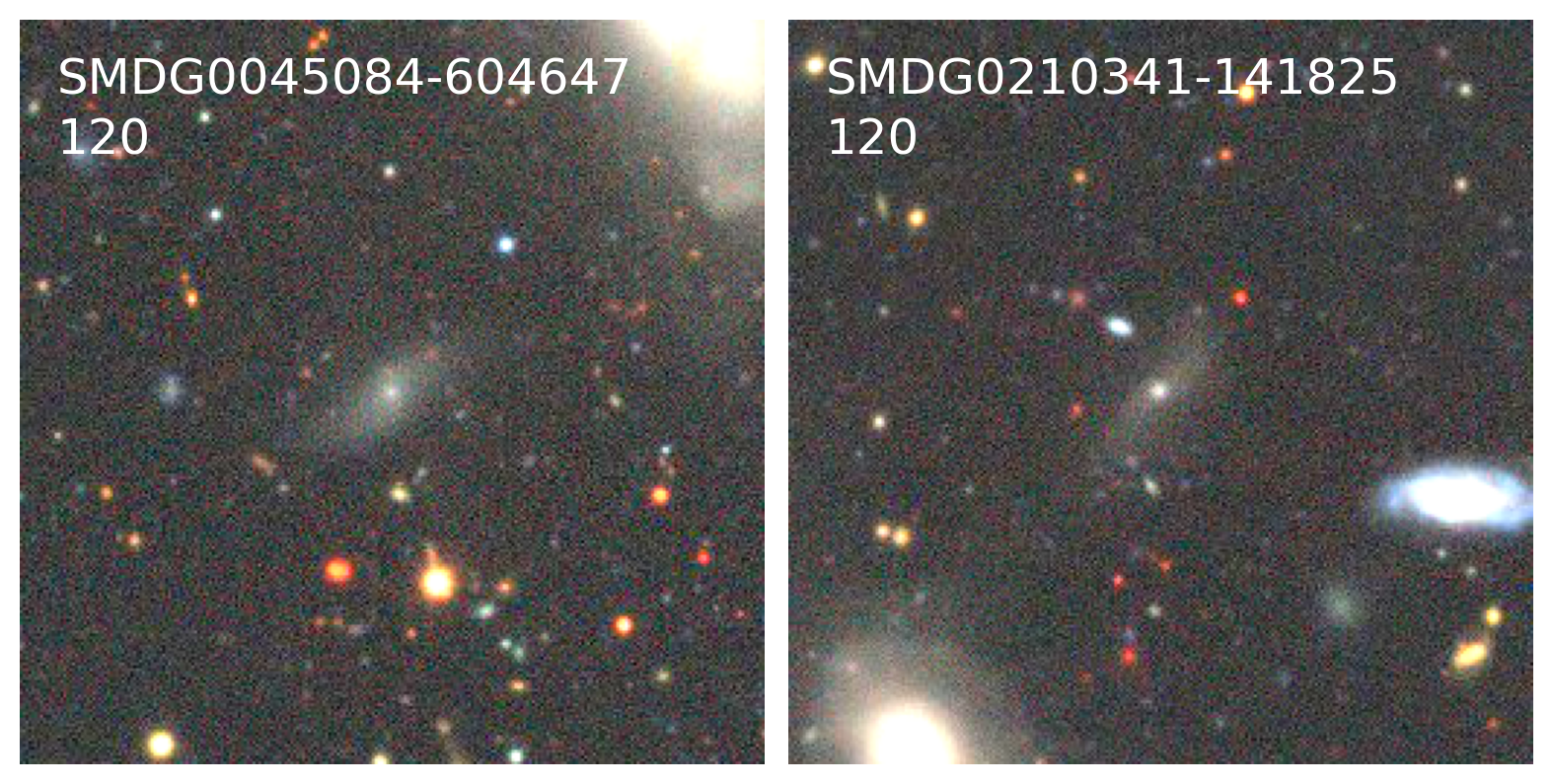}
\caption{Two examples of tidally distorted SMUDGes sources (Class = D). Images drawn from the Legacy Survey. The galaxy name and the size of the image in arcsec is given in each frame.}
\label{fig:tidally_distorted}
\end{figure}

In this class we place those galaxies that appear to have been galaxies prior to the recent, ongoing interaction but whose current appearance we interpret to be under the influence of tidal forces. In some cases, such as those in Figure \ref{fig:tidally_distorted} where there is a clear nucleus and the shape is consistent with the presence of tidal tails, the assignment is relatively straightforward. Not all cases are so clear. If there is no nucleus and if the galaxy is closer to the larger host galaxy, then it can be quite difficult to differentiate this class of galaxy from the next class that we will be discussing.
We classify 56 SMUDGes as tidally distorted and set them to have Class $=$ D for distorted. All of our classification labels are summarized in Table 2.

\begin{deluxetable}{lrr}
\label{tab:classes}
\tablewidth{0pt}
\tablecaption{Classification Labels}
\tablehead{\colhead{Category} & \colhead{Label}&
\colhead{\%}}
\startdata
Artifact&A&0.11\\
Background (generic)&B&0.61\\
Background binary&BB&0.38\\
Background group&BG&0.35\\
Background tidal&BT&0.52\\
Cirrus&C&1.15\\
Duplicate&DU&0.01\\
Emission line region&E&0.03\\
UDG candidate&G&92.69\\
Outer disk&OD&0.10\\
Ongoing merger&OM&0.11\\
Post merger&PM&0.10\\
Ring galaxy&R&0.41\\
Star& S&0.11\\
Smoke ring galaxy&SR&0.03\\
Tidal debris&T&2.22\\
Tidal distortion&D&0.79\\
Tidal dwarf&TDW&0.14\\
Tidal streamer&TS&0.13
\enddata
\end{deluxetable}

These galaxies may represent the transition phase in one highly plausible formation pathway. 
Interactions such as those experienced by this set of SMUDGs have been proposed as the origin story of at least some UDGs \citep{Carleton+2019,Liao+2019}. As such, we advocate for the retention of these in the sample of UDG candidates, keeping in mind that our cataloged parameters based on a single S\'ersic model are highly questionable. The fraction of such systems is modest ($\sim 57/7070)$, but 
UDG samples that favor dense environments may have a larger fraction of such systems. 

Evidence for a physical association with a more luminous host can be used to estimate a line-of-sight distance. We have a spectroscopic redshift for only one SMUDGe in this class, but a redshift is available from either SDSS or DESI for 25 identified host galaxies. Among these, only three SMUDGes have an estimated redshift obtained using the technique described by \cite{Zaritsky+2022}, where we associated a SMUDGe with a projected overdensity that is itself localized in redshift space. For two of the three, the estimated redshifts agree well. Given the limited number of objects for comparison, this level of agreement is consistent with the previously estimated accuracy rate for our estimated redshifts of 75\%.

\subsubsection{The Tidally Induced}
\label{sec:tidaldwarf}

Tidal forces can strip gas and stars from the outskirts of normal galaxies, and this material can concentrate to form putative dwarf galaxies \citep{zwicky,mirabel,duc,elmegreen,hunsberger}. Such
objects are also often referred to as tidal dwarfs, a classification class that precedes that of UDG by decades. A key physically distinguishing characteristic of tidal dwarfs is that they are expected to be relatively free of dark matter \citep{barnes}, and therefore fundamentally different from most galaxies. Some UDGs may be indeed be tidal dwarfs \citep{bennet}. Attempts have been made to confirm that hypothesis by establishing that these galaxies lack globular clusters \citep{jones}, which is another expected hallmark of galaxies that form from tidal material.

One challenge in accurately classifying such galaxies lies in confirming a tidal connection between an LSB satellite galaxy and its parent. Without clear bridging tidal material, one might instead identify the galaxy as an independent satellite. As such, it is possible that a non-negligible fraction of the SMUDGes with massive companions are tidal dwarfs. Although we have measured that for every two L$^*$ galaxies there is roughly one UDG satellite \citep{goto}, the fraction of those that are tidal dwarfs is unknown.

A second challenge in accurately classifying these galaxies is that it is difficult to assess whether a low surface brightness enhancement is truly a gravitationally bound system that will survive. For those that survive, we must accept that they 
differ in ways that are foundational for UDGs. Much of the current interest in UDGs arose from certain objects, e.g., DF 44 \citep{vanDokkum+2016}, because they are both dark matter dominated (to a level only seen for classical dwarf spheroidal and ultra-faint galaxies), and host a large number of globular clusters. Tidal UDGs would be counterexamples in both respects. It is then simply a coincidence of observational properties (stellar surface brightness and effective radius) that these two very different types of galaxy are considered together. We should not expect any single UDG formation mechanism to explain both categories.

We classify 10 SMUDGes as tidal dwarf galaxies and assign the Class $=$ TDW. This number represents $\sim$ 0.14\% of the overall catalog, which may suggest that the tidal dwarf UDG population is negligible. However, as already mentioned, these ten are only those with a clear tidal connection to the parent galaxy. Others may have a connection that is not bright enough for us to detect with our current images. Furthermore, those tidal UDGs that have survived for a sufficient time to move away from the vicinity of associated tidal debris will go undetected under our classification criteria, although one might expect many to fall back toward their parent and be either accreted or tidally disrupted. Finally, it is also possible that some systems with strong tidal features were rejected by our pipeline and excluded from the catalog. We conclude that there is yet little evidence to suggest that this is a common formation pathway, but we cannot reach a conclusion on what fraction of the overall population they comprise. None of the 10 we identify has a spectroscopic redshift measurement, but spectroscopic redshifts are available for nine of the ten parent galaxies. In all but two of the cases, the inferred $r_e$ is $>$ 7 kpc. Because these values are unusually large even for UDGs, the results suggest that the majority of these are not true galaxies. If these turn out to be gravitationally bound galaxies, then their measured $r_e$ values are likely to be highly inflated due to their current appearance.

\begin{figure}
\includegraphics[width=0.482\textwidth]{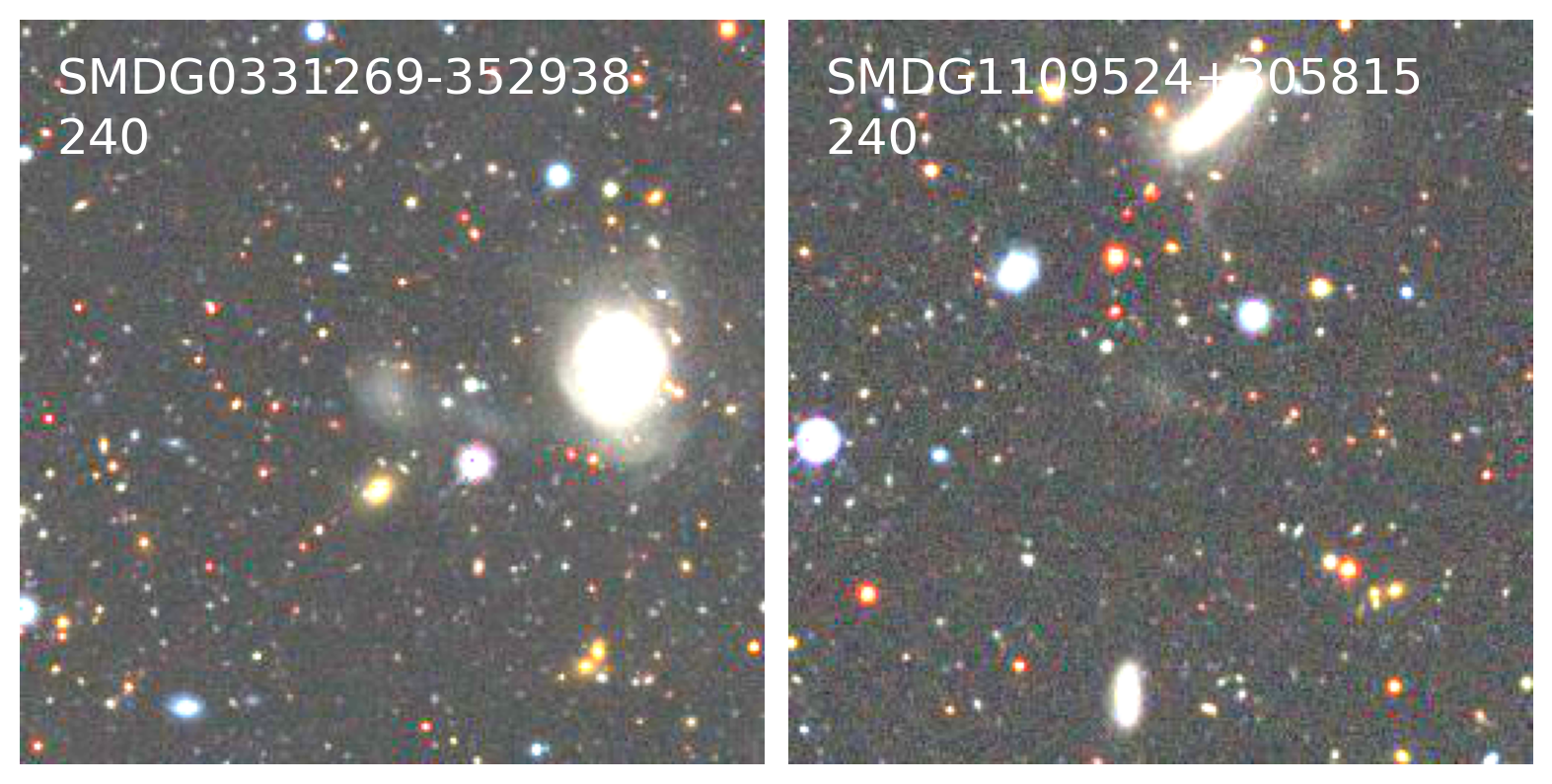}
\caption{Two potential examples of tidally induced SMUDGes sources (Class = TDW). Images drawn from the Legacy Survey.  The galaxy name and the size of the image in arcsec is given in each frame.}
\label{fig:tidally_induced}
\end{figure}

\subsubsection{LSB Galaxy Mergers}
\label{sec:lopsided}

The projected surface density of SMUDGes candidates is roughly 1/sq. degree. At this low projected density, it is unlikely to have two SMUDGes candidates adjacent to each other in projection, but galaxy clustering being what it is, we actually do find cases that appear to be mergers of LSB galaxies, perhaps even of UDGs. These examples may highlight a pathway to the formation of galaxies that are otherwise difficult to form in isolation, such as the largest UDGs, and may comprise unique systems with which to explore the dark matter properties of UDGs.

\begin{figure*}
\includegraphics[width=1\textwidth]{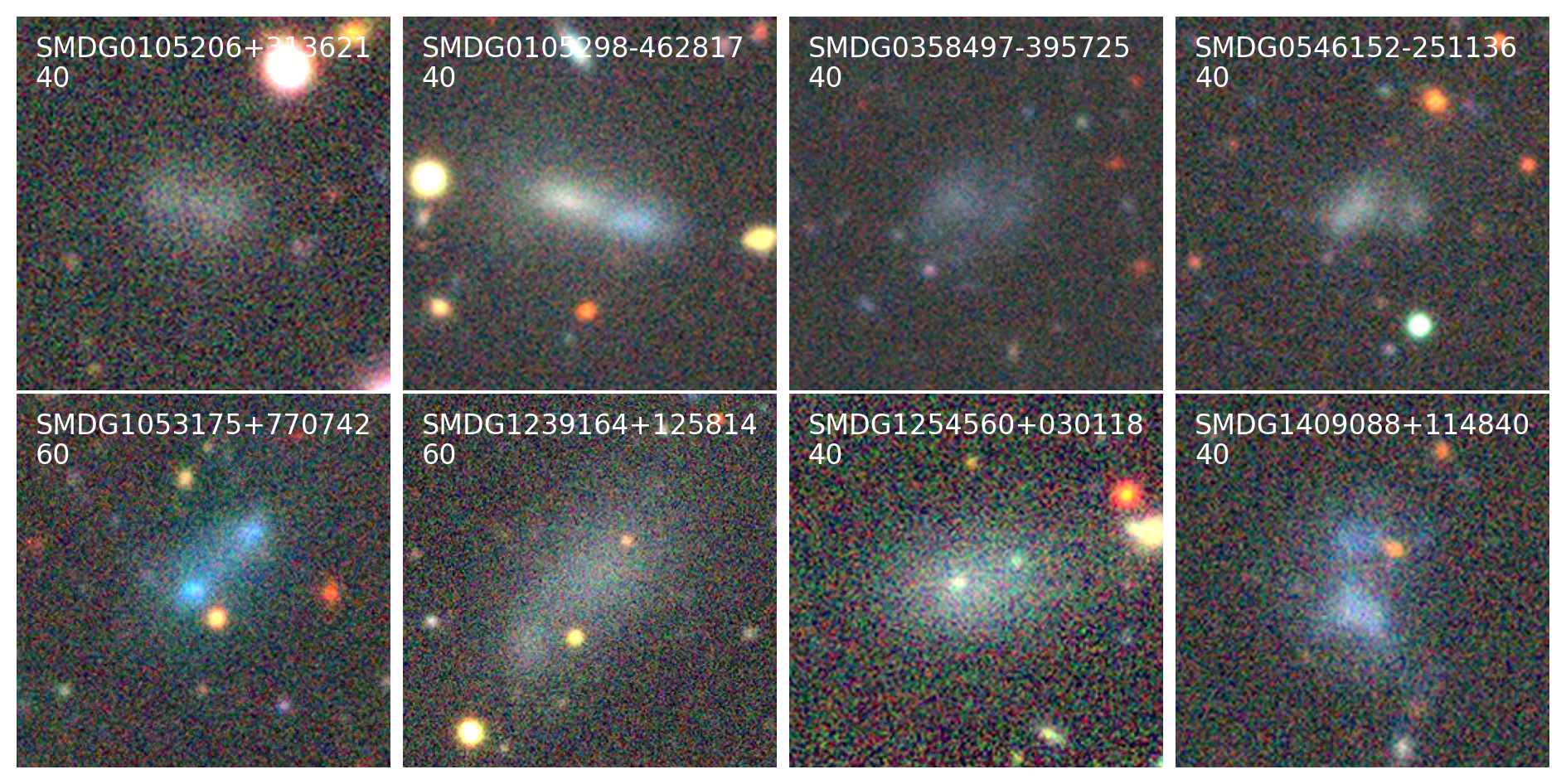}
\caption{Eight SMUDGes identified as ongoing mergers galaxies (Class = OM). Images drawn from the Legacy Survey.  The galaxy name and the size of the image in arcsec is given in each frame.}
\label{fig:dumbbell}
\end{figure*}

Our first subcategory in this class consists of systems that appear to have two cores and share a common, low surface brightness envelope (Figure \ref{fig:dumbbell}). For statistical reasons, these are likely to be physical pairs, although not necessarily as close as they appear in projection. If they are physically close, then perhaps they are the progenitors of the merged systems we will be discussing next. 
Among more massive galaxies that share this general appearance, i.e., `dumb-bell' galaxies, theoretical considerations suggest a rapid merger of the two components \citep{rix}. If these systems have the dark matter halos that appear to be generally present in galaxies, then we expect that dynamical friction will lead to a merger. Alternatively, some of these systems may be dwarf ring galaxies (see \S\ref{sec:rings}) that we are viewing edge on. 

Although it is difficult to reach a decisive conclusion on their nature without spectroscopy,
the case for at least some of these systems being physically-close, ongoing mergers rests on the tidal bridges that are visible in some (SMDG0546152-251136, SMDG1053175+770742, SMDG1239164+125814). The different colors of the two components in one case (SMDG0105298-462817) and the off-center nuclei on a couple of cases (SMDG1053175+770742, SMDG1254560+030118) argue against the edge-on ring galaxy scenario, at least in those cases.  We classify these as ongoing mergers and set Class = OM.

\begin{figure}
\includegraphics[width=0.472\textwidth]{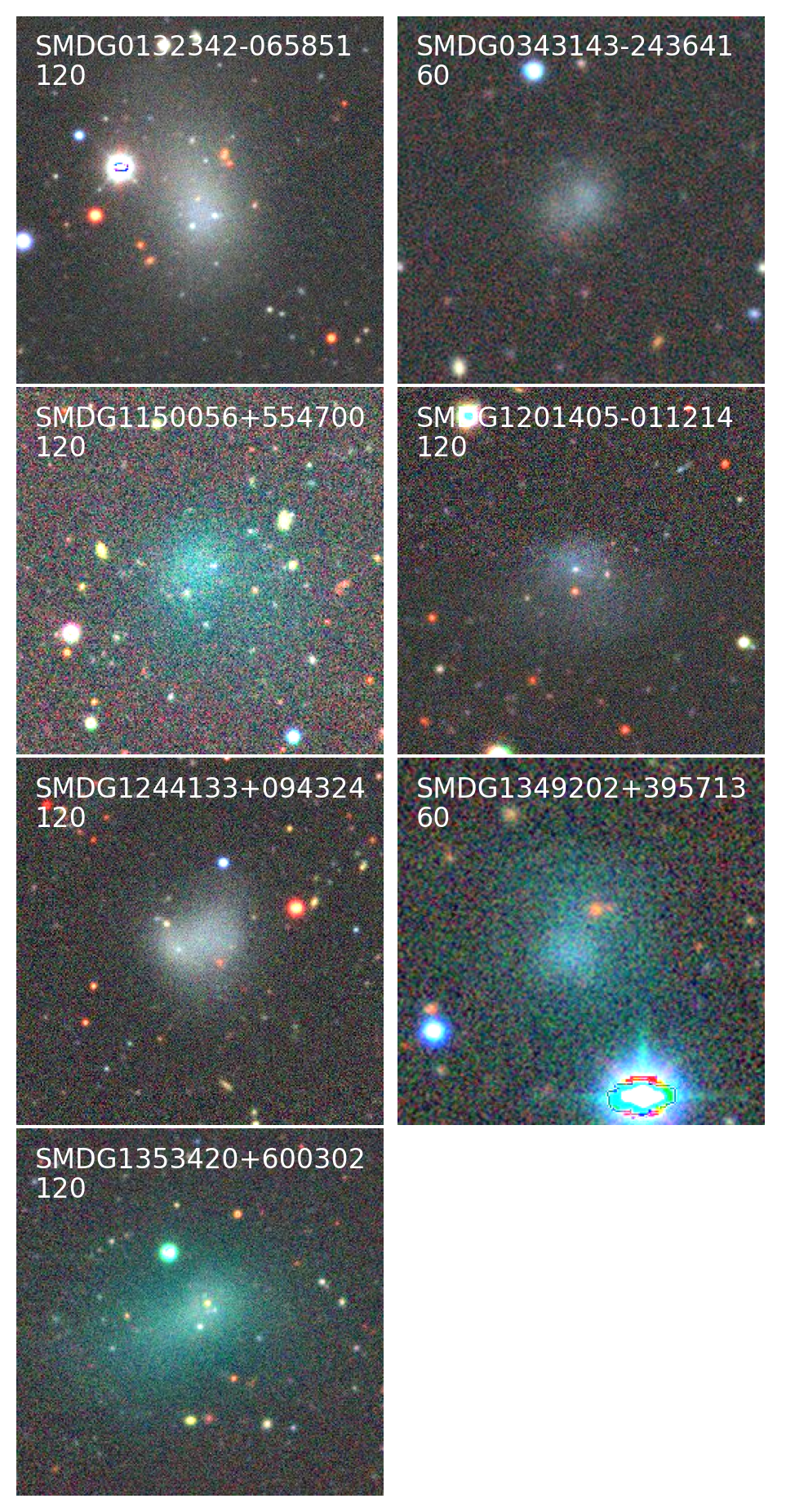}
\caption{The seven SMUDGes sources that are classified as post-merger (Class $=$ PM). Images drawn from the Legacy Survey.  The galaxy name and the size of the image in arcsec is given in each frame. The greenish shade of the sources in some of the panels is due to a color calibration issue in the creation of these images within this portion of the Legacy survey.}
\label{fig:lopsided}
\end{figure}

Further support for a merger scenario comes from our second subcategory, consisting of SMUDGes that appear to be post-merger systems (Class = PM). 
In Figure \ref{fig:lopsided} we show the seven systems we have classified as such. These are highly asymmetric UDG candidates, four of which perhaps have a nuclear star cluster. Although not evident from the Figure, most of these are isolated with no clear nearby perturber. Alternatively, the apparent lopsidedness could be due to internal structure such as a non-axisymmetric potential \citep{jog09}, but in at least two of the cases (SMDG1201405-011214 and SMDG1352420+600302) there are sharp surface brightness features reminiscent of shells, which are a hallmark of merger or accretion events. 

Once again, the concern for these objects is not that they are not LSB galaxies, but that the parameters as measured using a single S\'ersic model are not accurately representative. Interestingly, if these are mergers, they are mainly (5 of 7) red ($g-r> 0.55)$. There appears to have been no recent ($<$ few hundred Myr) star formation, suggesting that the progenitors did not have significant gas reservoirs. As part of future work, we will be quantifying the lopsidedness distribution among SMUDGes and deriving more quantitative estimates of the recent star formation histories to compare to the degree of morphological disturbance. For an example of a gas-rich UDG merger, which leads to star formation, see \cite{buzzo}. The SMUDGes survey may be biased against such systems if the episode of star formation raises the $g-$band central surface brightness above 24 mag arcsec$^{-2}$.

\begin{figure*}
\includegraphics[width=1\textwidth]{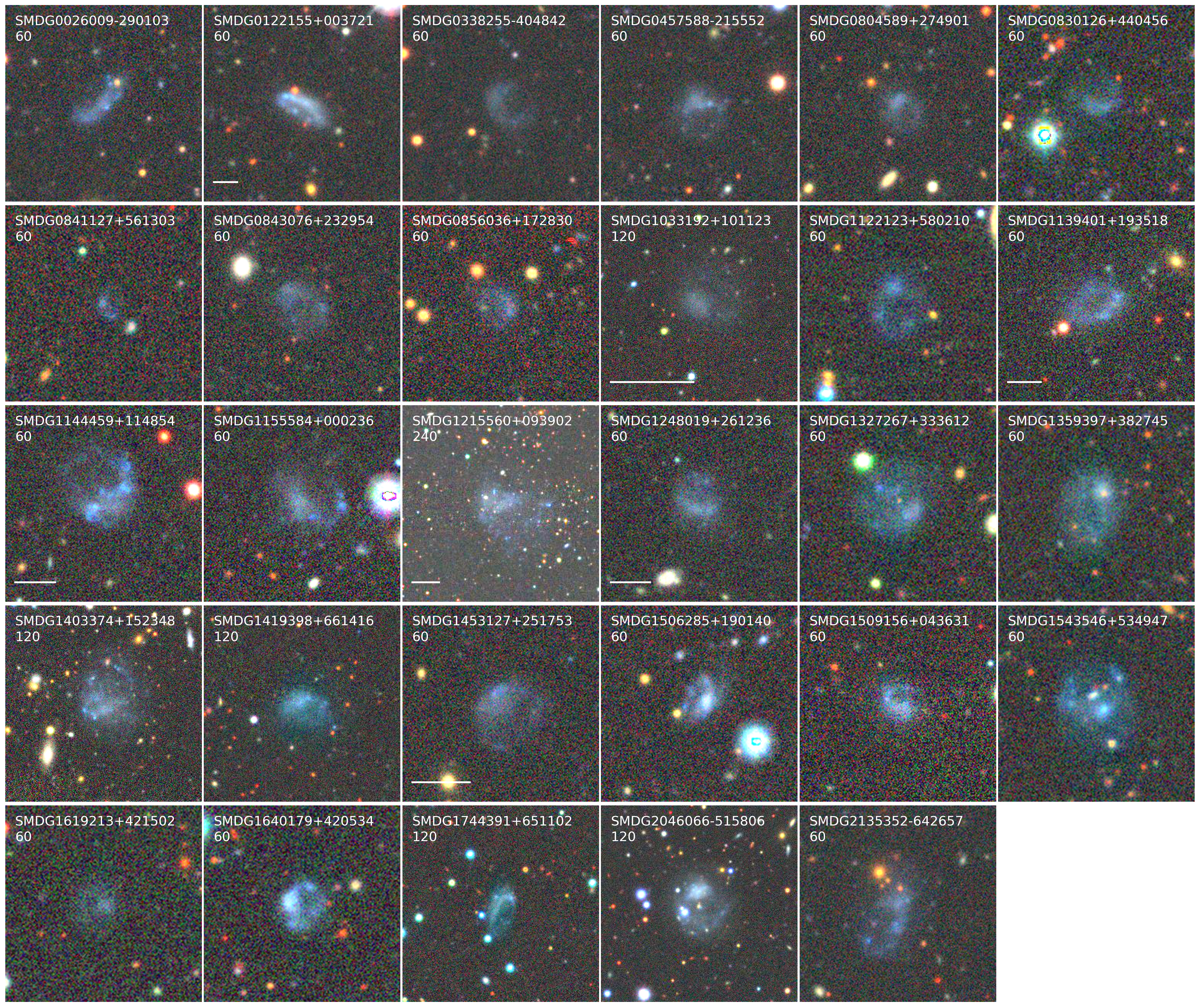}
\caption{The 29 SMUDGes identified as dwarf ring galaxies (Class = R). Images drawn from the Legacy Survey.  The galaxy name and the size of the image in arcsec is given in each frame. For those with available spectroscopic redshifts we had added a 5 kpc horizontal bar for scale in the lower left.}
\label{fig:rings}
\end{figure*}

\subsection{Miscasts}

\medskip

In this category, we place SMUDGes that are likely to be actual galaxies but whose path to becoming a UDG is unclear. Although a star-forming irregular galaxy may appear quite different from the canonical UDG, DF 44, one can envision how, once star formation stops, it may evolve to have a similar appearance. The objects that we place in this category do not have a clear pathway to appearing as a single S\'ersic model, although they might satisfy the observational central surface and size criteria that define UDGs.

\subsubsection{Dwarf Ring Galaxies}
\label{sec:rings}

An intriguing subclass of SMUDGes sources is that of the star-forming ring galaxy (Figure \ref{fig:rings}) and they are assigned Class $=$ R. These are blue, $\langle g-r \rangle = 0.32$ mag, circular galaxies with a central surface brightness depression. The rings are generally asymmetric with a large (resolved) bright region on one side and often, but not always, a gap in the ring across from that overdensity.
We have classified 29 SMUDGes as ring galaxies. Although we do have a few that appear to be highly inclined versions (see Figure \ref{fig:rings}), there is an observational bias against assigning Class $=$ R if the galaxy is highly inclined or at a large distance. As such, the actual number of such galaxies in the catalog is likely somewhat larger.

The SMUDGes pipeline usually identified the localized overdensity within the ring as the candidate UDG. The cataloged coordinates are therefore generally not at what one might more convincingly label the galaxy center. This is evident from the apparent miscentering of some objects in Figure \ref{fig:rings}. As such, the derived S\'ersic parameters are more likely to reflect the size and structure of the overdensity rather than of the entire galaxy, again requiring caution when using the cataloged parameters for these sources. Even so, for the seven galaxies with existing spectroscopic measurements, six have an inferred $r_e > 1.5$ kpc. The surface brightness is ill-defined for such irregular galaxies and the completeness estimates presented in the catalog, based on the injection of smooth objects into the survey data, are likely incorrect as well.

The formation of these objects is potentially interesting. Ring galaxies, in the case of more massive examples \citep{cannon,fosbury,madore}, are explained as the result of a nearly head-on collision (along the disk's polar axis) with a small, compact companion \citep{lynds}. To explain the observed morphology of our galaxies, we note that
the resulting rings produced by such collisions are expected to fragment \citep{inoue}, which appears to be common in our sample, and off-center collisions (which are more likely) produce asymmetric rings \citep{mapelli}.

With the SMUDGes being both fainter and of lower surface brightness than the ring galaxies considered previously, the responsible companions would be expected to be even smaller and fainter than those invoked for the more massive ring galaxies. Examining the images, we find no clearly responsible companions. If a collision is indeed the explanation for these objects and if the perturber survived, then the perturbing companion must be quite dim, suggesting a halo with a low stellar to halo mass ratio. If the former, such objects are of great interest in defining the low stellar mass end of the galaxy luminosity function \citep[for a review, see][]{bullock}.
One can envision reaching a point where the ring galaxy is of sufficiently low mass that the perturber could be a dark subhalo, below a mass where halos host significant stellar populations. Simulations should be carried out to determine the mass range of perturbers that could produce the observed rings, and follow-up spectroscopy should be carried out to assess possible perturbers.

\subsubsection{Smoke Rings}

\begin{figure}
\includegraphics[width=0.482\textwidth]{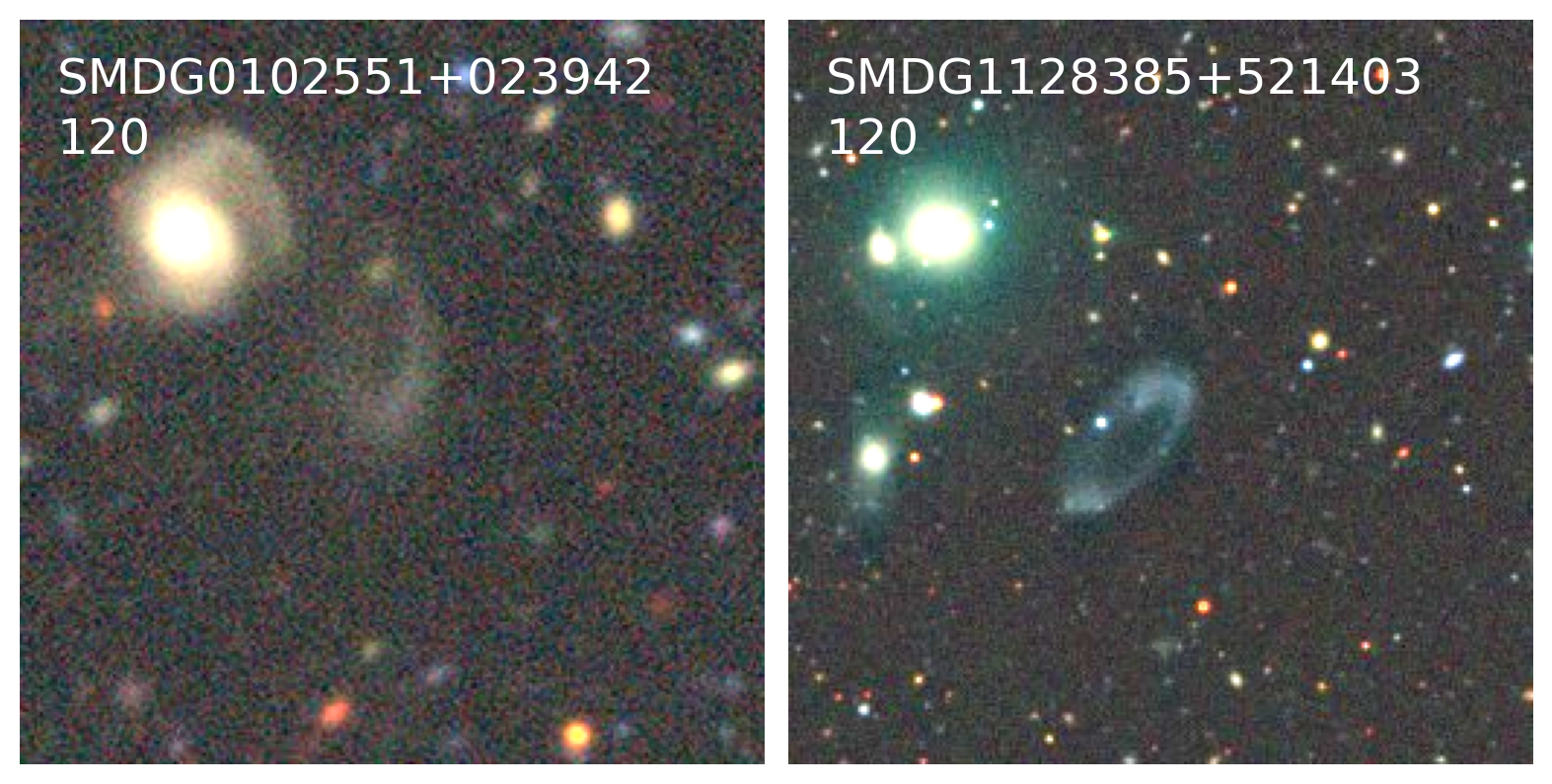}
\caption{The two SMUDGes identified as 'smoke rings' (Class = SR). Images drawn from the Legacy Survey.  The galaxy name and the size of the image in arcsec is given in each frame. These are off-center in the frame because the SMUDGes pipeline identified the brightest section along the ring as the galactic center.}
\label{fig:smokerings}
\end{figure}

A subclass of the ring galaxies consists of the two  SMUDGes shown in Figure \ref{fig:smokerings} (assigned Class $=$ SR and not included among the 29 ring galaxies discussed previously). These do not exhibit strong signs of star formation. The apparent greenish color of SMDG1128385+521403 in the Figure reflects a color calibration issue with the display in this portion of the Legacy survey, although it is the bluer of the two with $g-r = 0.47$ mag. The other galaxy has $g-r = 0.75$ mag. The key  distinguishing characteristic of this subclass is that these have a large, close companion that shows signs of tidal interaction. 
Despite their ring appearance, we interpret these as tidal features rather than stand-alone galaxies, which is why we present these separately.

\subsection{Miscreants}

\medskip

In this category we place detected objects that are masquerading as low luminosity galaxies. 

\subsection{Background Sources}

Among SMUDGes we reject are those that are galaxies, or sometimes multiple galaxies, that we classify as being well in the background and therefore not true low surface brightness dwarfs. We define several subclasses of these including a general background classification, Class $=$ B, binary galaxies that have been merged into a single detection, Class $=$ BB, tidal interactions among distant galaxies, Class $=$ BT, and groups of distant galaxies, Class $=$ BG. We classify 132 sources into one of these background categories.  Additional background sources may fall into some of the classes discussed below. Unfortunately, only seven among these have a measured spectroscopic redshift but their values do confirm that they lie significantly farther than the bulk of our objects (these seven all have $cz > 16,000$ km s$^{-1}$, with an average value of $\sim$ 61,000 km s$^{-1}$).

\subsubsection{Tidal Detritus}

Sometimes there is a fine line between a tidal dwarf (\S\ref{sec:tidaldwarf}) and an overdensity of loose tidal material seen in projection as a concentration. We classify 157 SMUDGes as tidal debris, class $=$ T. If we include the background galaxies with tidal features that we classify as Class $=$ BT, then the number increases to 194. 

We separate a distinct version of this class that results in a trail of star forming material (Class = TS). The distinction appears to be the result of whether the stripped material is principally stellar or gaseous. In the latter case, star formation occurs in the tidal tail, leading to a blue tidal streamer. We identify 9 such systems and present two examples in Figure {\ref{fig:streamers}.

\begin{figure}
\includegraphics[width=0.482\textwidth]{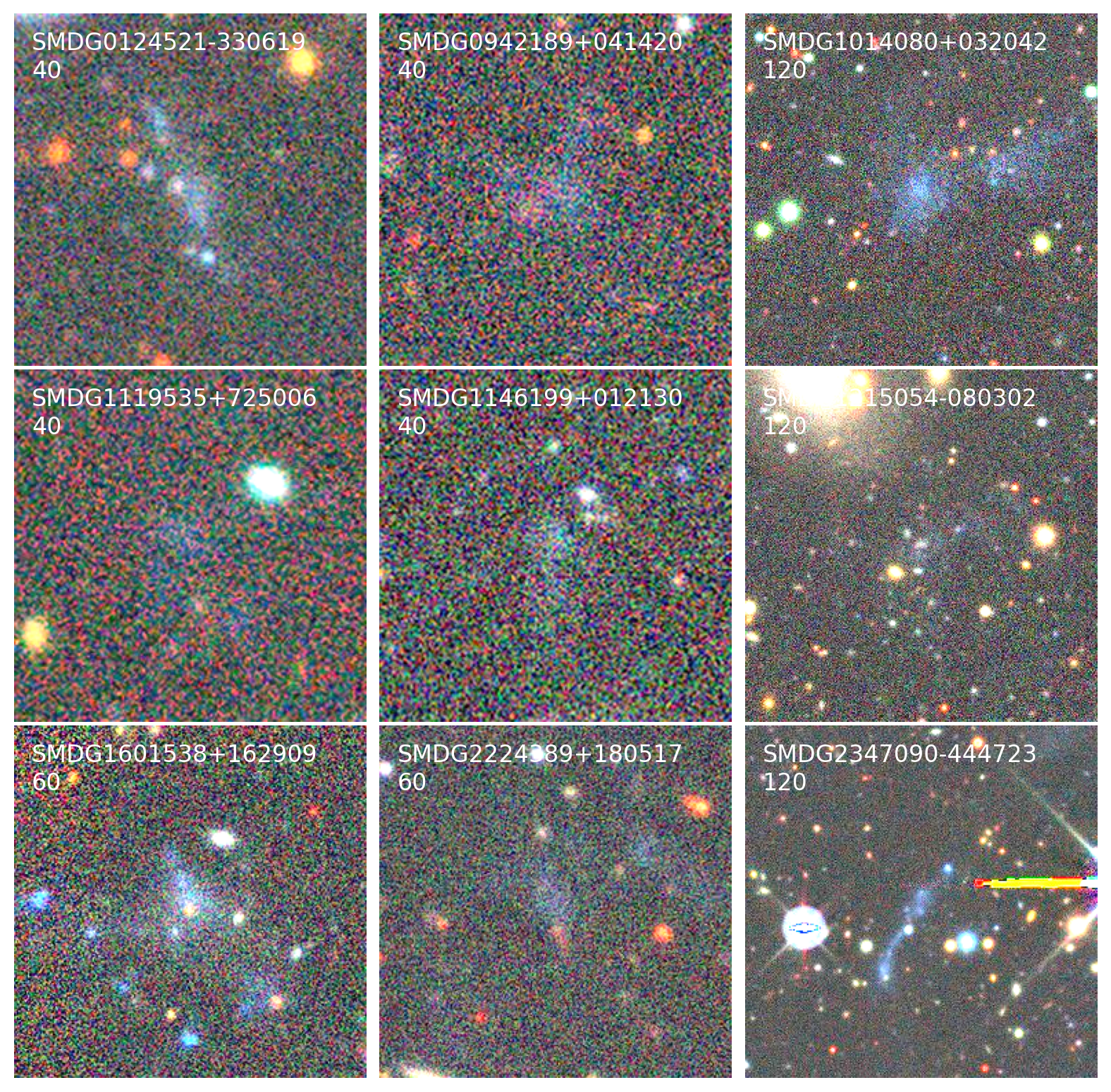}
\caption{All identified tidal streamers (Class = TS). These vary in morphology but generally consist of a filamentary distribution of distinct blue pockets of star formation. Images drawn from the Legacy Survey. The galaxy name and the size of the image in arcsec is given in each frame.}
\label{fig:streamers}
\end{figure}

\subsubsection{Cirrus}

We detailed previously \citep{Zaritsky+2021} our approach to removing most detections arising from localized Galactic Cirrus. Despite these efforts to reject Cirrus, as well as our application of a neural net classifier \citep{Zaritsky+2022}, we visually classify 81 SMUDGes as Cirrus and assign Class $=$ C.
None of these, as expected, have a spectroscopic redshift available.

\subsection{Remainders}

In this final category, we combine several classes, all of which should be rejected from further consideration as UDG candidates. First, there are 8 SMUDGes that appear to be artifacts, such as diffraction spikes or reflections. These are set to Class $=$ A, for artifact. Second, there are 7 SMUDGes that are overdensities within the outskirts of luminous, nearby star forming galaxies. We set Class $=$ OD for these outer disk sources. Third, there are 8 SMUDGes that are bright stars, and we set Class $=$ S for these. Fourth, there are 2 SMUDGes that are Galactic emission line regions along jets. Both of these sources are part of previously known sources.
The first (SMDG0333056-255813) is part of NGC 1360, a planetary nebula \citep{goldman}. The second (SMDG2322299-021040) is associated with a Herbig-Haro object \citep{zuckerman}. We set Class $=$ E for these objects.  
Finally, we set one duplicate detection to Class $=$ DU (the other detection is retained as a viable UDG candidate, Class = G).

\section{Discussion}
\label{sec:discussion}

Our exercise of re-visiting the SMDUGes catalog serves two purposes. First, it results in a higher-fidelity catalog of UDG candidates. The culling of the remaining sources associated with Galactic cirrus, background objects, artifacts, and stars that made it through our initial process is necessary. Second, it results in the identification of two interesting subclasses of SMUDGes that merit further scrutiny.

The first of those subclasses includes the SMUDGes labeled as ongoing mergers (OM) and post-mergers (PM). The number of such objects in the catalog is small (15), but interactions between dwarf galaxies could explain unusual systems \citep[e.g.,][]{bullet}. To those one could add a set of close pairs of SMUDGes objects to study the full merger sequence.  Although it becomes less certain to claim a physical association as the separation between UDG candidates increases, we do see a statistical excess of systems at small separations when we examine the distribution of pair separations (Figure \ref{fig:separations}), demonstrating that there are physical pairs in the sample. 
We calculate the results presented in the Figure by tabulating pair separations for candidates of Class = G, excluding candidates that we have previously identified to lie within the projected areas of the Coma, Fornax, and Virgo galaxy clusters. These results are meant only to substantiate the claim that at least some of these are physical close pairs, but a more thorough study is needed for quantitative results. In particular, the SMUDGes survey is roughly 50\% complete in large part due to significant masking near bright objects. If those masks are randomly distributed relative to the SMUDGes detections, then we would expect no systematic effect on the correlation function. However, \cite{goto} showed that  UDGs are often satellites of L$^*$ galaxies, and therefore we do expect masks to correlate with at least a fraction of the SMUDGes sources. 

\begin{figure}
\includegraphics[width=0.482\textwidth]{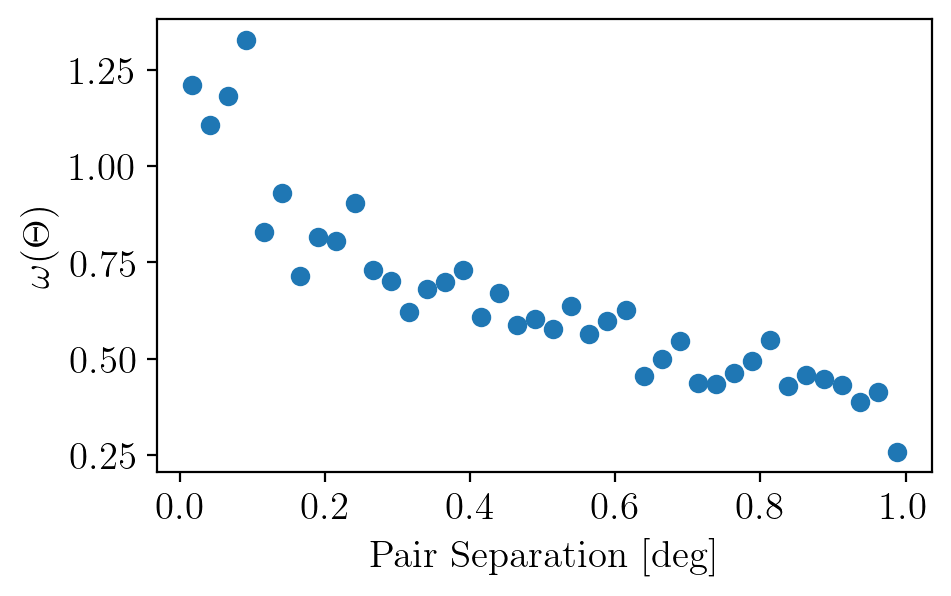}
\caption{Angular correlation function for SMUDGes candidates, which represents the excess fractional probability over random of finding a projected companion  as a function of the pair separation. We select only those objects of class = G and exclude those within the Coma, Fornax, and Virgo clusters as designated in \cite{smudges5}. There is no correction for masked survey regions. The rise toward small angular separations demonstrates an increasing statistical excess of close projected companions.}
\label{fig:separations}
\end{figure}

A large sample of isolated binary UDGs would be particularly valuable as it could be used to measure UDG halo masses to larger radii than currently possible. Inferences of UDGs lying in massive halos \citep[e.g.,][]{vanDokkum+2016,zaritsky} are based on the extrapolation of kinematic measurements at small radii ($\sim r_e$) to the virial radius using assumptions regarding a universal halo mass profile. A constraint at a radius $\gg r_e$ would address concerns about that extrapolation. We envision such a study along the lines of what has been done for massive binary galaxies \citep{white,white83}. Those studies suggest that meaningful results can be obtained with roughly a hundred pairs. 
Closer scrutiny of the SMUDGes pairs is necessary to assess how many are sufficiently isolated for such a study. That would  then be followed by the challenge of measuring the recessional velocities of a large sample of LSB galaxies.

The second of those subclasses is the dwarf ring galaxy. A kinematic study of these might prove fruitful in that many appear to host star-forming regions, easing the measurement of recessional velocities and internal kinematics. Simulations are needed to assess the types of encounter that can produce such structures in low-mass systems, whether the perturbers are likely to survive, and how long the rings survive as such. Answers to these questions would address viability and guide any search for the responsible perturber. Detailed simulations could assess whether dark subhalos are viable potential perturbers.

\section{Summary}
\label{sec:summary}
We have reviewed all 7,070 candidate ultra-diffuse galaxies (UDGs) in the SMUDGes catalog \citep{smudges5} with an eye toward identifying 1) those candidates, which while viable, are likely to have measured properties that are distorted by their interaction with nearby neighbors, 2) those candidates that might prove to be interesting examples of physical processes experienced by low surface brightness galaxies, and 3) those candidates that are unlikely to be UDGs. The last category includes detections that are galaxies and those that are not. Our classifications are presented in a new line-matched table to the original catalog. The number of objects in the catalog that are not flagged is 6,553.

Among the flagged objects, in the second of the three categories just mentioned, are a set of ongoing mergers between low surface brightness galaxies and a set of post-merger systems.
We have also identified a set of low surface brightness galaxies that appear to have had an interaction with a lower mass compact galaxy. 
All of these merit follow-up observation.

Catalogs always contain a subset of intriguing objects that might not match the original intent of the catalog and motivate further study. In this case, both higher angular resolution imaging and spectroscopy are likely to be critical in developing our understanding. Further scrutiny of the unflagged objects, as possible with deeper imaging, is likely to result in more flagged SMUDGes. We expect a review of the classifications in the next few years with the availability of {\sl Euclid}, Rubin / LSST and {\sl Roman} imaging.

\acknowledgments
\label{sec:acknowledgments}

\vspace{5mm}
\noindent
{\bf Software: }{\texttt{matplotlib \citep{matplotlib}, numpy \citep{numpy}, Astropy \citep{astropy1}, SciPy \citep{scipy1}}}

\bibliography{refs}{}
\bibliographystyle{aasjournal}

\end{document}